\documentclass[preprint,authoryear,12pt]{elsarticle}
\journal{International Journal of Plasticity}
\usepackage{graphics,epsfig}
\usepackage{graphicx}
\usepackage{amssymb,amstext,amsmath}
\usepackage{booktabs}
\usepackage{natbib}
\usepackage[latin1]{inputenc}
\usepackage{epstopdf}
\newcommand{\gta}{\mathfrak{a}}
\newcommand{\bfbeta}{\boldsymbol{\beta}}
\newcommand{\bfnu}{\boldsymbol{\nu}}
\newcommand{\bftau}{\boldsymbol{\tau}}
\newcommand{\bfvarepsilon}{\boldsymbol{\varepsilon}}
\newcommand{\bfsigma}{\boldsymbol{\sigma}}
\newcommand{\bfxi}{\boldsymbol{\xi}}
\newcommand{\tr}{\text{tr}}
\newcommand{\sign}{\text{sign}}
\begin{document}
\begin{frontmatter}
\title{Three-dimensional continuum dislocation theory}
\author{K. C. Le\footnote{phone: +49 234 32-26033, email: chau.le@rub.de. The paper is dedicated to the 70th birthday of my teacher V. Berdichevsky.}}
\address{Lehrstuhl f\"{u}r Mechanik - Materialtheorie, Ruhr-Universit\"{a}t Bochum,\\D-44780 Bochum, Germany}
\begin{abstract} 
A three-dimensional continuum dislocation theory for single crystals containing curved dislocations is proposed. A set of governing equations and boundary conditions is derived for the true placement, plastic slips, and loop functions in equilibrium that minimize the free energy of crystal among all admissible functions, provided the resistance to the dislocation motion is negligible. For the non-vanishing resistance to dislocation motion the governing equations are derived from the variational equation that includes the dissipation function. A simplified theory for small strains is also provided. An asymptotic solution is found for the two-dimensional problem of a single crystal beam deforming in single slip and simple shear.
\end{abstract}
\begin{keyword}
dislocations (A) \sep crystal plasticity (B) \sep finite strain (B) \sep variational calculus (C) .
\end{keyword}

\end{frontmatter}

\section{Introduction}\label{sec:introduction}

In view of a huge number of dislocations appearing in plastically deformed crystals (which typically lies in the range $10^8\div 10^{15}$ dislocations per square meter) the necessity of developing a physically meaningful continuum dislocation theory (CDT) to describe the evolution of dislocation network and predict the formation of microstructure in terms of mechanical and thermal loading conditions becomes clear to all researchers in crystal plasticity. One of the main guiding principles in seeking such a continuum dislocation theory has first been proposed by \citet{Hansen1986} in form of the so-called LEDS-hypothesis: the true dislocation structure in the final state of deformation minimizes the energy of crystal among all admissible dislocation configurations. In view of numerous experimental evidences supporting this hypothesis (see, i.e., \citep{Hughes1997,Kuhlmann1989,Kuhlmann2001,Laird1986}), its use in constructing the continuum dislocation theory seems to be quite reasonable and appealing. For the practical realization one needs to i) specify the whole set of unknown functions and state variables of the continuum dislocation theory, and ii) lay down the free energy of crystals as their functional to be minimized. Such program has been implemented by \citet{Berdichevsky06a} in the linear, and by \citet{Le2014nonlinear} in the nonlinear setting of CDT for networks of dislocations, whose lines are straight and remain so during the whole deformation process (see also \citep{Ortiz99,Ortiz00}). The developed CDT has been successfully applied to various two-dimensional problems of dislocation pileups, bending, torsion, as well as formation of dislocation patterns in single crystals (see \citep{Berdichevsky-Le07,Kaluza2011torsion,Kochmann08b,Kochmann08a,Kochmann09,Koster2015,Le08a,Le08b,Le2009plane,Le2012polygonization,Le2013on}). Let us mention here the similar approaches suggested in \citep{Acharya2000,Acharya2001,Engels2012,Gurtin2002,Gurtin2007,Mayeur2014,Oztop2013} which do not use the LEDS-hypothesis explicitly but employ instead the extended principle of virtual work for the gradient plasticity. However, as experiences and experiments show, dislocation lines are in general loops that, as a rule, can change their directions and curvatures depending on the loading condition and crystal's geometry. Therefore the extension of CDT to  networks of dislocations whose lines are curves in the slip planes is inevitable. To the best of author's knowledge, such three-dimensional continuum dislocation theory based on the LEDS-hypothesis for curved dislocations has not been developed until now. It became also clear to him that the latter's absence was due to the missing scalar dislocation densities for the network of curved dislocations. 

The first attempt at constructing a continuum theory that can predict in principle not only the dislocation densities but also the direction and curvature of the dislocation lines has been made by \citet{Hochrainer2007} in form of the so-called continuum dislocation dynamics. Their theory starts with the definition of the dislocation density that contains also the information about the orientation and curvature of the dislocation lines. Then the set of kinematic equations is derived for the dislocation density and curvature that requires the knowledge about the dislocation velocity. The relation between the dislocation density and the macroscopic plastic slip rate via the dislocation velocity is postulated in form of Orowan's equation. The couple system of crystal plasticity and continuum dislocation dynamics becomes closed by the constitutive equation of a flow rule type (see \citep{Hochrainer2014,Sandfeld2011,Sandfeld2015,Wulfinghoff2015}). In addition to the heavy computational cost of such theory, the relation to thermodynamics of crystal plasticity and to the LEDS-hypothesis is completely lost: the equilibrium solution found in this theory may not minimize the energy of crystal among all admissible dislocation configurations. Let us mention also a continuum approach proposed recently by \citet{Zhu2013,Zhu2014} in which the three-dimensional dislocation structure is characterized by two families of disregistry functions that may take only integer values. The dislocation density can then be expressed in their terms. The coupled system of equations is derived from the underlying discrete dislocation dynamics for the displacement and disregistry functions. This approach is subject to the same critics as that proposed in \citep{Hochrainer2007}.

The aim of this paper is to extend the nonlinear continuum dislocation theory (CDT) developed recently by \citet{Le2014nonlinear} to the case of crystals containing curved dislocations. Provided the dislocation network is regular in the sense that nearby dislocations have nearly the same direction and orientation, we introduce a loop function whose level curves coincide with the dislocation lines. Taking an infinitesimal area perpendicular to the dislocation line at some point of the crystal, we express the densities of edge and screw dislocations at that point through the resultant Burgers vectors of dislocations whose lines cross this area at right angle. Such scalar densities contain not only the information about the number of dislocations, but also the information about the orientation and curvature of the dislocation lines. In case of dislocation motion we introduce the vector of normal velocity of dislocation line through the time derivative of the loop function. Following \citet{Kroener92} and \citep{Berdichevsky06} we require that the free energy density of crystal depends only on the elastic strain tensor and on the above scalar densities of dislocations. Then we formulate a new variational principle of CDT according to which the placement, the plastic slip, and the loop function in the final state of equilibrium minimize the free energy functional among all admissible functions. We derive from this variational principle a new set of equilibrium equations, boundary conditions, and constitutive equations for these unknown functions. In case the resistance to dislocation motion is significant, the variational principle must be replaced by the variational equation that takes the dissipation into account. The constructed theory is generalized for single crystals having a finite number of active slip systems. We provide also the simplifications of the theory for small strains. As compared to the continuum dislocation dynamics proposed in \citep{Hochrainer2007,Zhu2014} our theory is advantageous not only in the computational cost due to its simplicity, but also in its full consistency with the LEDS-hypothesis. In the problem of single crystal beam having only one active slip system and deforming in simple shear, the energy minimization problem reduces to the two-dimensional variational problem. We solve this problem analytically for the circular cross section and asymptotically for the rectangular cross section. We will show that this solution reduces to that found in \citep{Berdichevsky-Le07} for the crystals with thin and long cross-section.   

The paper is organized as follows.  After this short introduction we present in Section 2 the three-dimensional kinematics for single crystals deforming in single slip. Section 3 formulates the variational principles of the three-dimensional CDT and derives its governing equations. Section 4 extends this nonlinear theory to the case of single crystals with $n$ active slip system. Section 5 studies the three-dimensional small strain CDT. Section 6 is devoted to the analytical and asymptotic solutions of the two-dimensional energy minimization problem of a single crystal beam deforming in simple shear. Finally, Section 7 concludes the paper.

\section{3-D kinematics for single crystals deforming in single slip}\label{sec:ncdt}

\begin{figure}[htb]
	\centering
	\includegraphics[height=6.5cm]{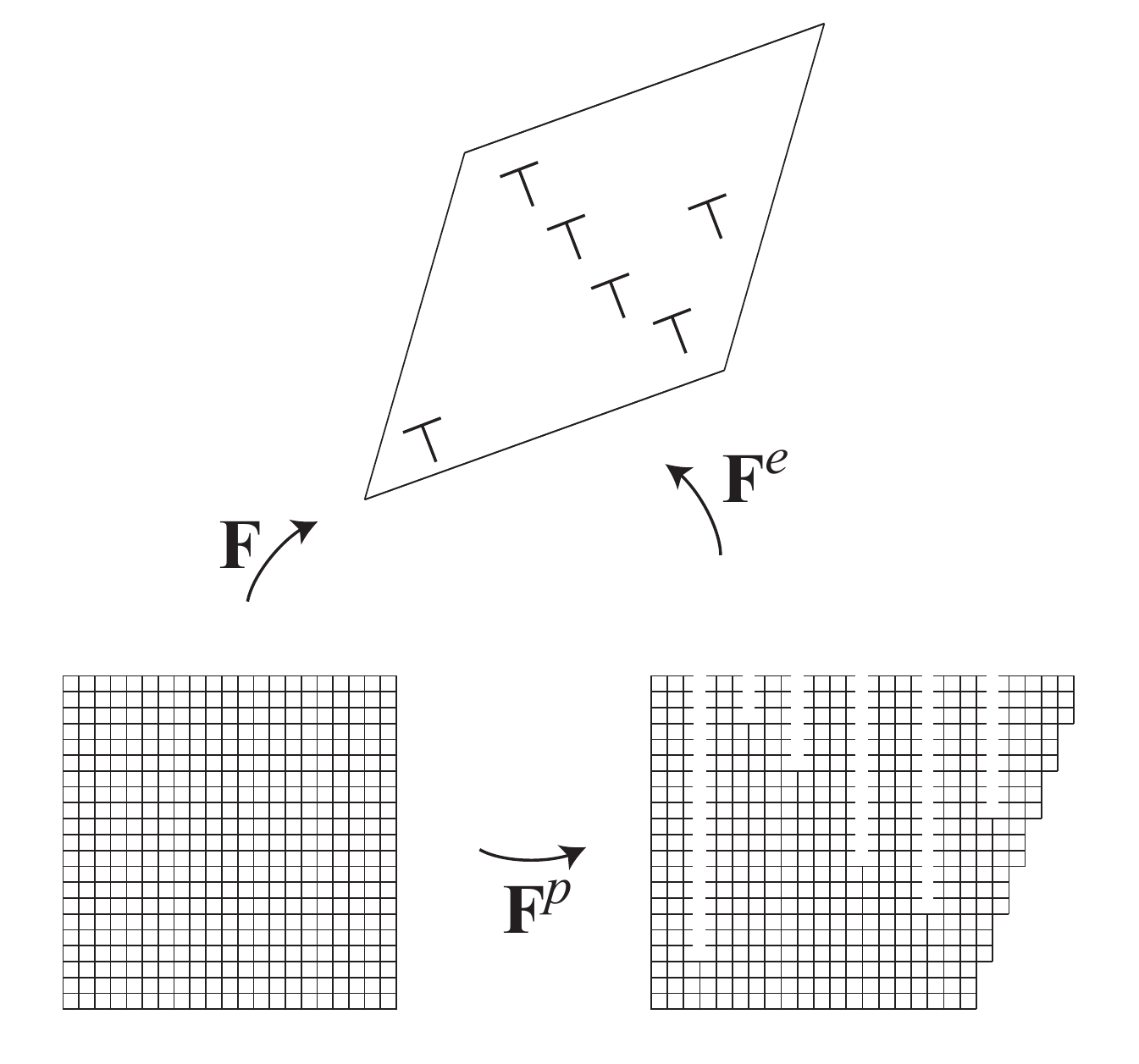}
	\caption{Multiplicative decomposition}
	\label{fig:Nhslip}
\end{figure}

Nonlinear CDT starts from the basic kinematic resolution of the deformation gradient $\mathbf{F}=\partial \mathbf{y}/\partial \mathbf{x}$ into elastic and plastic parts \citep{Bilby57} 
\begin{equation}\label{eq:2.1}
	\mathbf{F}=\mathbf{F}^e\cdot \mathbf{F}^p.
\end{equation}
We attribute an active role to the plastic deformation: $\mathbf{F}^p$ is the deformation {\it creating} dislocations (either inside or at the boundary of the volume element) or {\it changing} their positions in the crystal without distorting the lattice parallelism (see Fig.~\ref{fig:Nhslip}). On the contrary, the elastic deformation $\mathbf{F}^e$ deforms the crystal lattice having {\it frozen} dislocations \citep{Le2014nonlinear}. Note that the lattice vectors remain unchanged when the plastic deformation is applied, while they change together with the shape vectors by the elastic deformation.

\begin{figure}[htb]
	\centering
	\includegraphics[height=6.5cm]{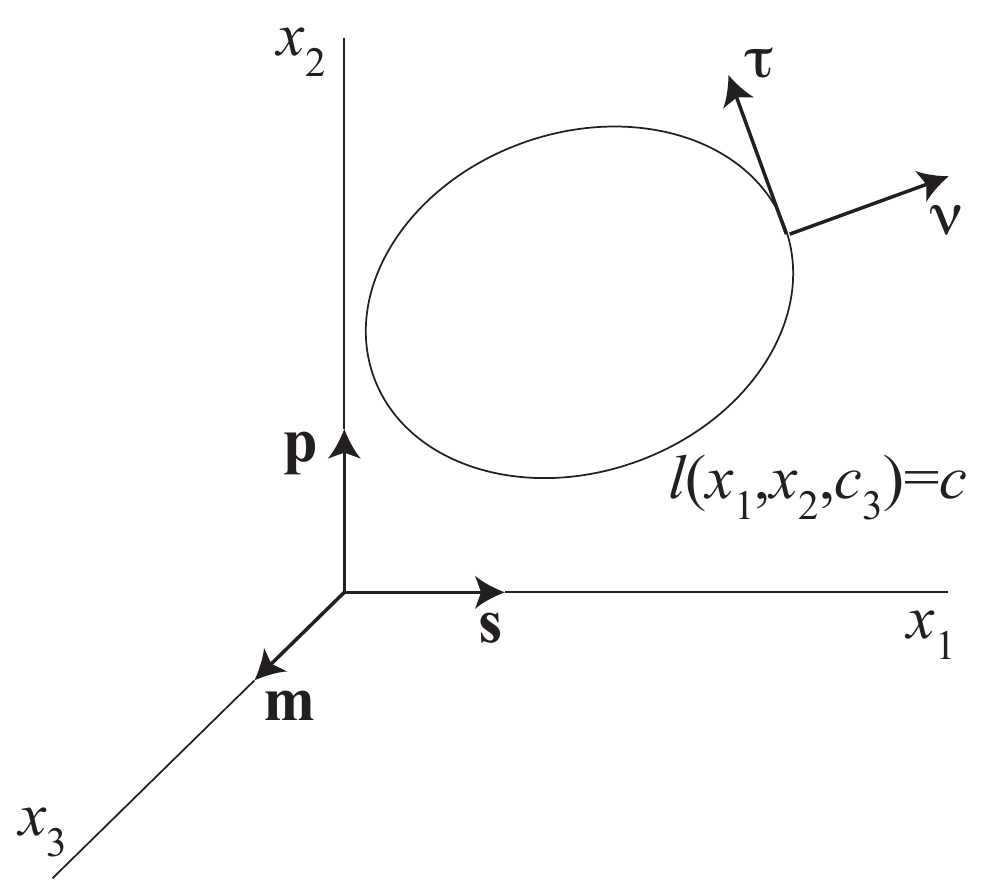}
	\caption{A dislocation loop in the chosen coordinate system}
	\label{fig:loop}
\end{figure}

We consider first a single crystal deforming in single slip. In this case let us denote the right-handed triad of unit lattice vectors of the active slip system by $\mathbf{s}$, $\mathbf{p}$, and $\mathbf{m}$, where $\mathbf{s}$ points to the slip direction, $\mathbf{p}$ lies in the slip plane and is perpendicular to $\mathbf{s}$, and $\mathbf{m}$ is normal to the slip plane. Without restricting generality we may choose the rectangular cartesian coordinate system $(x_1,x_2,x_3)$ in the reference configuration such that its basis vectors coincide with these lattice vectors (see Fig.~\ref{fig:loop})
\begin{equation*}
\mathbf{e}_1=\mathbf{s},\quad \mathbf{e}_2=\mathbf{p}, \quad \mathbf{e}_3=\mathbf{m}.
\end{equation*}
The plastic deformation is then given by
\begin{equation}
\label{eq:2.0}
\mathbf{F}^p=\mathbf{I}+\beta (\mathbf{x})\mathbf{s}\otimes \mathbf{m}=\mathbf{I}+\beta (\mathbf{x})\mathbf{e}_1\otimes \mathbf{e}_3,
\end{equation}
with $\beta $ being the plastic slip. We assume that all dislocations causing this plastic deformation lie completely in the slip planes and the dislocation network is regular in the sense that nearby dislocations have nearly the same direction and orientation. This enables one to introduce a scalar function $l(x_1,x_2,x_3)$ (called a loop function) such that its level curves 
\begin{equation}
\label{eq:2.a}
l(x_1,x_2,c_3)=c,
\end{equation} 
with $c_3$ and $c$ being constants, coincide with the dislocation lines. Thus, in this three-dimensional kinematics we admit, according to equation \eqref{eq:2.a}, only the conservative motion of dislocations and exclude from consideration the dislocation climb which is an important mechanism of temperature-dependent creep. We denote by $\bfnu $ and $\bftau $ the plane unit vectors normal and tangential to the dislocation line. From equation \eqref{eq:2.a} follow
\begin{equation*}
\bfnu =\frac{1}{\sqrt{l_{,1}^2+l_{,2}^2}}(l_{,1}\mathbf{e}_1+l_{,2}\mathbf{e}_2), \quad \bftau =\frac{1}{\sqrt{l_{,1}^2+l_{,2}^2}}(-l_{,2}\mathbf{e}_1+l_{,1}\mathbf{e}_2),
\end{equation*}
where the comma before an index denotes the partial derivative with respect to the corresponding coordinate. Note that $\bfnu $, $\bftau $, $\mathbf{m}$ form a right-handed basis vectors of the three-dimensional space (see Fig.~\ref{fig:loop}). 

\citet{Ortiz99} introduced the resultant Burgers vector of excess dislocations, whose lines cross the area $\mathcal{A}$ in the reference configuration, in the following way
\begin{equation}\label{eq:2.2}
\mathbf{b}_r=\oint_{\mathcal{C}} \mathbf{F}^p\cdot d\mathbf{x} , 
\end{equation}
where $\mathcal{C}$ is the close contour surrounding $\mathcal{A}$. \citet{Le2014nonlinear} have shown that, in the continuum limit, when the atomic distance goes to zero at the fixed sizes of the representative volume element and the fixed density of dislocations per area of unit cell, integral \eqref{eq:2.2} gives the total closure failure induced by $\mathbf{F}^p$ which must be equal to the resultant Burgers vector. It is natural  to assume $\mathbf{F}^p$ continuously differentiable in this continuum limit, so, applying Stoke's theorem we get from \eqref{eq:2.2}
\begin{equation*}
\mathbf{b}_r=- \int _{\mathcal{A}}(\mathbf{F}^p\times \nabla ) \cdot \mathbf{n}da,
\end{equation*}
where $\times $ denotes the vector product, $da$ the surface element, and $\mathbf{n}$ the unit vector normal to $\mathcal{A}$. This legitimates the introduction of the dislocation density tensor
\begin{equation*}
\mathbf{T}=-\mathbf{F}^p\times \nabla .
\end{equation*}
For the plastic deformation taken from \eqref{eq:2.0} 
\begin{equation*}
\mathbf{T}=-\mathbf{F}^p\times \nabla =\mathbf{s}\otimes (\nabla \beta \times \mathbf{m}).
\end{equation*}
If we choose now an infinitesimal area $da$ with the unit normal vector $\bftau $, then the resultant Burgers vector of all excess dislocations, whose dislocation lines cross this area at right angle is given by
\begin{equation*}
\mathbf{b}_r=\mathbf{T}\cdot \bftau \, da=-\mathbf{s}\, (\nabla \beta \cdot \bfnu )da=-\mathbf{s}\, \partial _\nu \beta \, da.
\end{equation*}
This resultant Burgers vector can be decomposed into the sum of two vectors 
\begin{equation*}
\mathbf{b}_r=\mathbf{b}_{r\perp}+\mathbf{b}_{r\parallel}=-(\bfnu s_\nu +\bftau s_\tau )\partial _\nu \beta\, da,
\end{equation*}
where $s_\nu =\mathbf{s}\cdot \bfnu =\nu _1$ and $s_\tau =\mathbf{s}\cdot \bftau =\tau _1$ are the projections of the slip vector onto the normal and tangential direction to the dislocation line, respectively. This allows us to define two scalar densities (or the numbers of excess dislocations per unit area) of edge and screw dislocations 
\begin{equation}
\label{eq:2.3a}
\begin{split}
\rho_\perp =\frac{|\mathbf{b}_{r\perp}|}{b}=\frac{1}{b}|s_\nu \partial _\nu \beta| =\frac{1}{b}\left| \frac{l_{,1}(\beta _{,1}l_{,1}+\beta _{,2}l_{,2})}{l_{,1}^2+l_{,2}^2}\right| , 
\\
\rho_\parallel =\frac{|\mathbf{b}_{r\parallel}|}{b}=\frac{1}{b}|s_\tau \partial _\nu \beta|=\frac{1}{b}\left| \frac{l_{,2}(\beta _{,1}l_{,1}+\beta _{,2}l_{,2})}{l_{,1}^2+l_{,2}^2}\right| ,
\end{split}
\end{equation}
with $b$ the magnitude of Burgers vector. We see that the three-dimensional dislocation densities $\rho_\perp $ and $\rho_\parallel $ depend on both the gradient of the plastic slip and the gradient of the loop function $l(\mathbf{x})$ through the vectors $\bfnu $ and $\bftau $.

Consider now the case of motion of dislocation loops in the slip plane. In this case we allow the loop function to depend explicitly on time $t$ such that equation
\begin{equation}
\label{eq:2.4}
l(x_1,x_2,c_3,t)=c
\end{equation}
with fixed constants $c_3$ and $c$ describes one and the same dislocation line during its motion in the slip plane. Letting $\zeta $ be the variable along the dislocation line, we may represent the level curve defined by \eqref{eq:2.4} in the parametric form
\begin{equation*}
x_1=x_1(\zeta ,t),\quad x_2=x_2(\zeta ,t).
\end{equation*}
Fixing $\zeta $ and taking the differential of \eqref{eq:2.4} we obtain
\begin{equation*}
l_{,1}dx_1+l_{,2}dx_2+l_{,t}dt=0
\end{equation*}
that yields
\begin{equation*}
l_{,1}\frac{dx_1}{dt}+l_{,2}\frac{dx_2}{dt}=-l_{,t}.
\end{equation*}
Since $\mathbf{v}=\frac{dx_1}{dt}\mathbf{e}_1+\frac{dx_2}{dt}\mathbf{e}_2$ is the velocity of the fixed point on the dislocation line with coordinate $\zeta $, we define the normal velocity of the dislocation line as follows
\begin{equation}
\label{eq:2.5}
v_\nu =\mathbf{v}\cdot \bfnu =\frac{1}{\sqrt{l_{,1}^2+l_{,2}^2}}(l_{,1}\frac{dx_1}{dt}+l_{,2}\frac{dx_2}{dt})=-\frac{\dot{l}}{\sqrt{l_{,1}^2+l_{,2}^2}},
\end{equation}
with $\dot{l}=l_{,t}$. This kinematic quantity will be used in the case of non-vanishing resistance to dislocation motion.

\section{Governing equations for single crystals deforming in single slip}

According to \citet{Kroener92}, the elastic deformation $\mathbf{F}^e$ and the dislocation densities $\rho_\perp $ and $\rho_\parallel $ characterize the current state of the crystal, so these quantities are the state variables of the continuum dislocation theory. The reason why the plastic deformation $\mathbf{F}^p$ cannot be qualified for the state variable is that it depends on the cut surfaces and consequently on the whole  history of creating dislocations. Likewise, the gradient of plastic strain tensor $\mathbf{C}^p$ cannot be used as the state variable by the same reason. In contrary, the dislocation densities depend only on the characteristics of dislocations in the current state (Burgers vector and positions of dislocation lines) and not on how they are created, so $\rho_\perp $ and $\rho_\parallel $, in addition to $\mathbf{F}^e$, are the proper state variables. Thus, if we consider isothermal processes of deformation, then the free energy per unit volume of crystal (assumed as macroscopically homogeneous) must be a function of $\mathbf{F}^e$, $\rho_\perp $, and $\rho_\parallel $ 
\begin{equation*}
\psi = \psi (\mathbf{F}^e,\rho_\perp ,\rho_\parallel ).
\end{equation*}
Now, if we superimpose an elastic rotation $\mathbf{R}$ onto the actual deformation of the body, then the total and elastic deformation change according to 
\begin{equation*}
\mathbf{F}^{\ast}=\mathbf{R}\cdot \mathbf{F},\quad \mathbf{F}^{e\ast}=\mathbf{R}\cdot \mathbf{F}^e.
\end{equation*}
At the same time, the dislocation densities $\rho_\perp $ and $\rho_\parallel $ remain unchanged. As such superimposed elastic rotation does not change the elastic strain and the dislocation densities, we expect that the energy remains unchanged. The standard argument (see, e.g., \citep{Gurtin1981}) leads then to
\begin{equation*}
\psi = \psi (\mathbf{C}^e,\rho_\perp ,\rho_\parallel ),
\end{equation*}
where $\mathbf{C}^e$ is the elastic strain defined by
\begin{equation*}
\mathbf{C}^e=\mathbf{F}^{eT}\cdot \mathbf{F}^e.
\end{equation*}

Let the undeformed single crystal occupy some region $\mathcal{V}$ of the three-dimensional  space. The boundary of this region, $\partial \mathcal{V}$, is assumed to be the closure of union of two non-intersecting surfaces, $\partial _k$ and $\partial _s$. Let the placement be a given smooth function of coordinates
\begin{equation}\label{eq:2.6}
\mathbf{y}(\mathbf{x})=\mathbf{x}+\mathbf{u}_0(\mathbf{x}) \quad \text{at $\partial _k$},
\end{equation}
where $\mathbf{u}_0(\mathbf{x})=\mathbf{y}(\mathbf{x})-\mathbf{x}$ is the given displacement vector. Such condition does not admit dislocations to reach this part $\partial _k$ of the boundary, so we set 
\begin{equation}
\label{eq:2.6a}
\beta (\mathbf{x})=0, \quad l(\mathbf{x})=0 \quad \text{at $\partial _k$}.
\end{equation}
At the remaining part $\partial _s$ the ``dead'' load (traction) $\mathbf{t}$ is specified. 
Note that, in case the whole boundary is free, we do not have any kinematic constraint at $\partial \mathcal{V}$. However, as the energy density is invariant with respect to the shift of the loop function $l(\mathbf{x})$ on an arbitrary constant which does not change the dislocation densities, we can impose on this scalar function the following constraint
\begin{equation*}
\int_{\mathcal{V}}l(\mathbf{x})\, dx=0,
\end{equation*}
where $dx=dx_1dx_2dx_3$ denotes the volume element. If no body force acts on this crystal, then its energy functional is defined as
\begin{equation}
\label{eq:2.7}
I[\mathbf{y}(\mathbf{x}),\beta (\mathbf{x}),l(\mathbf{x})]=\int_{\mathcal{V}}w(\mathbf{F},\beta ,\nabla \beta ,\nabla l)\, dx-\int_{\partial _s} \mathbf{t} \cdot \mathbf{y}\, da,
\end{equation}
where
\begin{equation}\label{eq:2.8}
w(\mathbf{F},\beta ,\nabla \beta ,\nabla l)=\psi (\mathbf{C}^e,\rho_\perp ,\rho_\parallel ),
\end{equation}
Provided the resistance to the dislocation motion is negligibly small and no surfaces of discontinuity occur inside crystals, then the following variational principle is valid for single crystals with one active slip system: the true placement vector $\check{\mathbf{y}}(\mathbf{x})$, the true plastic slip $\check{\beta }(\mathbf{x})$, and the true loop function $\check{l}(\mathbf{x})$ in the {\it final} equilibrium state of deformation minimize energy functional \eqref{eq:2.7} among all continuously differentiable fields $\mathbf{y}(\mathbf{x})$, $\beta (\mathbf{x})$, and $l(\mathbf{x})$ satisfying constraints \eqref{eq:2.6} and \eqref{eq:2.6a}. 

Let us derive the equilibrium equations from this variational principle. We compute the first variation of functional \eqref{eq:2.7}
\begin{equation*}
\delta I= \int_{\mathcal{V}}
\left( \mathbf{P}\mathbf{:}\delta \mathbf{y}\nabla +\frac{\partial w}{\partial \beta }\delta \beta +\frac{\partial w}{\partial \nabla \beta }\cdot \nabla \delta \beta +\frac{\partial w}{\nabla l}\cdot \nabla \delta l \right) dx
- \int_{\partial _s} \mathbf{t} \cdot \delta \mathbf{y}\, da , 
\end{equation*}
where $\mathbf{P}=\partial w /\partial \mathbf{F}$. Integrating the first, third, and fourth term by parts with the help of Gauss' theorem and taking the conditions \eqref{eq:2.6} and \eqref{eq:2.6a} into account, we obtain
\begin{multline}\label{eq:2.9}
\delta I = \int_{\mathcal{V}}[ -\delta \mathbf{y}\cdot (\mathbf{P}\cdot \nabla ) +(w_\beta -\nabla \cdot w_{\nabla \beta })\delta \beta - (\nabla \cdot w_{\nabla l})\delta l ] \, dx
\\
+ \int_{\partial _s} [(\mathbf{P}\cdot \mathbf{n}-\mathbf{t} )\cdot \delta \mathbf{y} +w_{\nabla \beta }\cdot \mathbf{n}\, \delta \beta]\, da + \int_{\partial \mathcal{V}}w_{\nabla l }\cdot \mathbf{n}\, \delta l \, da=0.
\end{multline}
Equation \eqref{eq:2.9} implies that the minimizer must satisfy in $\mathcal{V}$ the equilibrium equations
\begin{equation}\label{eq:2.10}
\mathbf{P}\cdot \nabla =0,
\quad
-w_\beta +\nabla \cdot w_{\nabla \beta }=0, \quad \nabla \cdot w_{\nabla l} =0,
\end{equation} 
subjected to the kinematic boundary conditions \eqref{eq:2.6} and \eqref{eq:2.6a} at $\partial _k$, and the following natural boundary conditions 
\begin{equation}\label{eq:2.11}
\mathbf{P}\cdot \mathbf{n}=\mathbf{t},
\quad
w_{\nabla \beta }\cdot \mathbf{n}=0 , \quad w_{\nabla l }\cdot \mathbf{n}=0 \quad \text{at $\partial _s$}.
\end{equation}
We call $\mathbf{P}$ the first Piola-Kirchhoff stress tensor, $\tau _r=-w_\beta $ the resolved shear stress (or Schmid stress), and $\varsigma =-\nabla \cdot w_{\nabla \beta }$ the back stress. The first equation of \eqref{eq:2.10} is nothing else but the equilibrium of macro-forces acting on the crystal, the second equation represents the equilibrium of micro-forces acting on dislocations, while the last one expresses the equilibrium condition for the curved dislocation lines. 

The constitutive equations for $\mathbf{P}=w_{\mathbf{F}}$, $-w_\beta $, $w_{\nabla \beta }$, and $w_{\nabla l}$ can easily be obtained from the free energy density \eqref{eq:2.8}. First, we express $\mathbf{F}^e$ in terms of $\mathbf{F}$ and $\beta $ with the use of \eqref{eq:2.1} and \eqref{eq:2.0}
\begin{equation*}
\mathbf{F}^e=\mathbf{F}\cdot \mathbf{F}^{p-1}=\mathbf{F}\cdot (\mathbf{I}-\beta \mathbf{s}\otimes \mathbf{m}).
\end{equation*}
Now, the standard differentiation using the chain rule and the above relation yields the first Piola-Kirchhoff stress tensor
\begin{equation}\label{eq:2.12}
\mathbf{P}=w_{\mathbf{F}}=2\mathbf{F}^e\cdot \psi _{\mathbf{C}^e} \cdot \mathbf{F}^{p-T}.
\end{equation}
For the resolved shear stress (Schmid stress) we get
\begin{equation}\label{eq:2.13}
\tau _r=-w_\beta =2\mathbf{s}\cdot \mathbf{F}^{p-T}\cdot \mathbf{C}^e \cdot \psi _{\mathbf{C}^e}\cdot \mathbf{m}.
\end{equation}
Likewise, from \eqref{eq:2.3a} follows
\begin{equation}
\label{eq:2.14}
w_{\nabla \beta }=\frac{1}{b}[\psi _{\rho_\perp }\sign (s_\nu \partial _\nu \beta )s_\nu +\psi _{\rho_\parallel }\sign (s_\tau \partial _\nu \beta )s_\tau ] \bfnu.
\end{equation}
Thus, the vector $w_{\nabla \beta }$ is two-dimensional. Finally, we compute $w_{\nabla l }$ directly in components using formulas \eqref{eq:2.3a}. Since $\rho _\perp $ and $\rho _\parallel $ do not depend on $l_{,3}$, so $w_{l_{,3}}=0$, and the vector $w_{\nabla l}$ is also two-dimensional. For its first two components we have
\begin{equation}
\label{eq:2.15}
\begin{split}
w_{l_{,1}}=\frac{1}{b}\left[ \psi _{\rho _\perp }\sign (s_\nu \partial _\nu \beta )\left( -\frac{2l_{,1}^2(\beta _{,1}l_{,1}+\beta _{,2}l_{,2})}{(l_{,1}^2+l_{,2}^2)^2}+\frac{2\beta _{,1}l_{,1}+\beta _{,2}l_{,2}}{l_{,1}^2+l_{,2}^2} \right) \right.
\\
+\left. \psi _{\rho _\parallel }\sign (s_\tau \partial _\nu \beta )\left( -\frac{2l_{,1}l_{,2}(\beta _{,1}l_{,1}+\beta _{,2}l_{,2})}{(l_{,1}^2+l_{,2}^2)^2}+\frac{\beta _{,1}l_{,2}}{l_{,1}^2+l_{,2}^2} \right) \right] ,
\\
w_{l_{,2}}=\frac{1}{b}\left[ \psi _{\rho _\perp }\sign (s_\nu \partial _\nu \beta )\left( -\frac{2l_{,1}l_{,2}(\beta _{,1}l_{,1}+\beta _{,2}l_{,2})}{(l_{,1}^2+l_{,2}^2)^2}+\frac{\beta _{,2}l_{,1}}{l_{,1}^2+l_{,2}^2} \right) \right.
\\
+\left. \psi _{\rho _\parallel }\sign (s_\tau \partial _\nu \beta )\left( -\frac{2l_{,2}^2(\beta _{,1}l_{,1}+\beta _{,2}l_{,2})}{(l_{,1}^2+l_{,2}^2)^2}+\frac{\beta _{,1}l_{,1}+2\beta _{,2}l_{,2}}{l_{,1}^2+l_{,2}^2} \right) \right] .
\end{split}
\end{equation}
Substituting the constitutive equations \eqref{eq:2.12}-\eqref{eq:2.15} into \eqref{eq:2.10}-\eqref{eq:2.11} we get the completely new system of equations and boundary conditions which, together with \eqref{eq:2.6} and \eqref{eq:2.6a}, enable one to determine $\check{\mathbf{y}}(\mathbf{x})$, $\check{\beta }(\mathbf{x})$, and $\check{l}(\mathbf{x})$. Note that equations \eqref{eq:2.10}$_1$ and \eqref{eq:2.10}$_2$ are coupled via the first Piola-Kirchhoff stress tensor and the Schmid stress containing both $\mathbf{F}$ and $\beta $, while equations \eqref{eq:2.10}$_2$ and \eqref{eq:2.10}$_3$ are coupled because both contain the gradients of $\beta $ and $l$. All equations are strongly nonlinear partial differential equations.

The above theory has been developed for the case of negligibly small resistance to dislocation motion and plastic slip. In real crystals there is however always the resistance to the dislocation motion and plastic slip causing the energy dissipation that changes the above variational principle as well as the equilibrium conditions. We assume that the dissipation function depends on the plastic slip rate $\dot{\beta }$ and on the normal velocity of the dislocation loop $v_\nu $ given by \eqref{eq:2.5} (or, equivalently, on $\dot{l}$). Thus,
\begin{equation*}
D=D(\dot{\beta },\dot{l}),
\end{equation*}
When the dissipation is taken into account, the above formulated variational principle must be modified. Following \citep{Sedov65,Berdichevsky1967} we require that the true placement $\check{\mathbf{y}}(\mathbf{x},t)$, the true plastic slips $\check{\beta }(\mathbf{x},t)$, and the true loop function $\check{l}(\mathbf{x},t)$ obey the variational equation
\begin{equation}
\label{eq:2.15b}
\delta I+\int_{\mathcal{V}} (\frac{\partial D}{\partial \dot{\beta }}\delta \beta +\frac{\partial D}{\partial \dot{l}}\delta l)\, dx=0
\end{equation}
for all variations of admissible fields $\mathbf{y}(\mathbf{x},t)$, $\beta (\mathbf{x},t)$, and $l(\mathbf{x},t)$ satisfying the constraints \eqref{eq:2.6} and \eqref{eq:2.6a}. Together with the above formula for $\delta I$ and the arbitrariness of $\delta \mathbf{y}$, $\delta \beta $, and $\delta l$ in $\mathcal{V}$ as well as at $\partial _s$, equation \eqref{eq:2.15b} yields
\begin{equation}
\label{eq:2.15c}
\mathbf{P}\cdot \nabla =0,
\quad
-w_\beta +\nabla \cdot w_{\nabla \beta }=\frac{\partial D}{\partial \dot{\beta }}, \quad \nabla \cdot w_{\nabla l}=\frac{\partial D}{\partial \dot{l}},
\end{equation}
which are subjected to the kinematic boundary conditions \eqref{eq:2.6} and \eqref{eq:2.6a}, and the natural boundary conditions \eqref{eq:2.11}. The constitutive equations remain exactly the same as \eqref{eq:2.12}-\eqref{eq:2.15}. For the rate-independent theory the dissipation function can be assumed in a simple form
\begin{equation*}
D=K_1|\dot{\beta }|+K_2|\dot{l}|,
\end{equation*}
with $K_1$ and $K_2$ being positive constants. We call $K_1$ the critical resolved shear stress and $K_2$ the Peierls threshold. In this case equations \eqref{eq:2.15c}$_{2,3}$ become
\begin{equation*}
-w_\beta +\nabla \cdot w_{\nabla \beta }=K_1\, \sign \dot{\beta }, \quad \nabla \cdot w_{\nabla l}=K_2\, \sign \dot{l}
\end{equation*}
for non-vanishing $\dot{\beta}$ and $\dot{l}$. These are the yield conditions for $\beta $ and $l$: $\dot{\beta}$ and $\dot{l}$ are non-zero if and only if 
\begin{equation*}
|-w_\beta +\nabla \cdot w_{\nabla \beta }|=K_1, \quad |\nabla \cdot w_{\nabla l}|=K_2.
\end{equation*}
On the contrary, if the expressions on the left-hand sides are less than $K_1$ and $K_2$, the plastic slip cannot evolve and the dislocation lines cannot move: $\dot{\beta}=0$ and $\dot{l}=0$. Thus, they are frozen in the crystal.

\section{Extension to multiple slip}

The extension to the case of single crystals having $n$ active slip systems can be done straightforwardly under the assumption\footnote{In conventional crystal plasticity the kinematic equation for $\mathbf{F}^p$ is usually formulated in rate form that does not always reduces to \eqref{eq:2.16} \citep{Ortiz99,Ortiz00}.}
\begin{equation}
\label{eq:2.16}
\mathbf{F}^p=\mathbf{I}+\sum_{\mathfrak{a}=1}^n \beta ^\mathfrak{a}(\mathbf{x}) \mathbf{s}^\mathfrak{a}\otimes \mathbf{m}^\mathfrak{a} ,
\end{equation}
with $\beta ^\mathfrak{a}$ being the plastic slip, where the pair of constant and mutually orthogonal unit vectors $\mathbf{s}^\mathfrak{a}$ and $\mathbf{m}^\mathfrak{a}$ is used to denote the slip direction and the normal to the slip planes of the corresponding $\mathfrak{a}$-th slip system, respectively. Here and later, the Gothic upper index $\mathfrak{a}$ running from 1 to $n$ numerates the slip systems, so one could clearly distinguish $\beta ^\gta $ from the power function. We denote by $\mathbf{p}^\gta $ the unit vector lying in the slip plane such that $\mathbf{s}^\gta $, $\mathbf{p}^\gta $, and $\mathbf{m}^\gta $ form a right-handed basis vectors. For each slip system we can introduce the coordinates associated with these basis vectors
\begin{equation}
\label{eq:2.17}
\xi ^\gta _1=\mathbf{s}^\gta \cdot \mathbf{x}, \quad \xi ^\gta _2=\mathbf{p}^\gta \cdot \mathbf{x},\quad \xi ^\gta _3=\mathbf{m}^\gta \cdot \mathbf{x}.
\end{equation}
Equations \eqref{eq:2.17} can be regarded as the one-to-one linear transformation relating $\bfxi ^\gta $ and $\mathbf{x}$ according to
\begin{equation*}
\bfxi ^\gta =\mathbf{M}^\gta \cdot \mathbf{x}, \quad \mathbf{x}=\mathbf{M}^{\gta -1}\bfxi ^\gta ,
\end{equation*}
where $\mathbf{M}^\gta $ is the $3\times 3$ matrix whose rows are basis vectors $\mathbf{s}^\gta $, $\mathbf{p}^\gta $, and $\mathbf{m}^\gta $. Thus, any function of $\mathbf{x}$ can be expressed as function of $\bfxi ^\gta $ and {\it vice versa}. For the plastic slip $\beta ^\mathfrak{a}$ caused by dislocations of the slip system $\gta $ we assume that their lines lie completely in the slip planes parallel to the $(\xi^\gta _1,\xi^\gta _2)$-plane. To describe the latter we introduce the loop function $l^\gta (\xi^\gta _1,\xi^\gta _2, \xi^\gta _3)$ such that its level curves
\begin{equation}
\label{eq:2.18}
l^\gta (\xi ^\gta_1,\xi ^\gta _2,c _3)=c,
\end{equation} 
where $c_3$ and $c$ are constants, coincide with the dislocation lines. We denote by $\bfnu ^\gta $ and $\bftau ^\gta $ the plane unit vectors normal and tangential to the dislocation line. From equation \eqref{eq:2.18} follow
\begin{equation*}
\bfnu ^\gta =\frac{1}{\sqrt{(l^\gta _{;1})^2+(l^\gta _{;2})^2}}(l^\gta _{;1}\mathbf{s}^\gta +l^\gta _{;2}\mathbf{p}^\gta ), \quad \bftau ^\gta =\frac{1}{\sqrt{(l^\gta _{;1})^2+(l^\gta _{;2})^2}}(-l^\gta _{;2}\mathbf{s}^\gta +l^\gta _{;1}\mathbf{p}^\gta ),
\end{equation*}
where the semicolon in indices denotes the partial derivatives of the loop function with respect to $\xi ^\gta _1$, $\xi ^\gta _2$, so these vectors lie in the slip planes parallel to the $(\xi _1,\xi _2)$-plane as expected.

For the plastic deformation \eqref{eq:2.16} the dislocation density tensor becomes
\begin{equation}\label{eq:2.20}
\mathbf{T}=-\mathbf{F}^p\times \nabla =\sum_{\mathfrak{a}=1}^n \mathbf{s}^\gta \otimes (\nabla \beta ^\gta \times \mathbf{m}^\gta ).
\end{equation}
To characterize the geometrically necessary dislocations belonging to one slip system we consider one term $\mathbf{T}^\gta =\mathbf{s}^\gta \otimes (\nabla \beta ^\gta \times \mathbf{m}^\gta )$ in the sum \eqref{eq:2.20}. Let us choose an infinitesimal area $da$ with the unit normal vector $\bftau ^\gta $ and compute the resultant Burgers vector of all excess dislocations of the system $\gta $, whose dislocation lines cross this area at right angle 
\begin{equation*}
\mathbf{b}^\gta _r=\mathbf{T}^\gta \cdot \bftau ^\gta \, da=-\mathbf{s}^\gta \, (\nabla \beta ^\gta \cdot \bfnu ^\gta )da=-\mathbf{s}^\gta \, \partial _{\nu ^\gta }\beta ^\gta \, da.
\end{equation*}
This resultant Burgers vector can be decomposed into the sum of two vectors 
\begin{equation*}
\mathbf{b}^\gta _r=\mathbf{b}^\gta _{r\perp}+\mathbf{b}^\gta _{r\parallel}=-(\bfnu ^\gta s^\gta _\nu +\bftau ^\gta s^\gta _\tau )\partial _{\nu ^\gta }\beta^\gta \, da,
\end{equation*}
where $s^\gta _\nu =\mathbf{s}^\gta \cdot \bfnu ^\gta $ and $s^\gta _\tau =\mathbf{s}^\gta \cdot \bftau ^\gta $ are the projections of the slip vector onto the normal and tangential direction to the dislocation line, respectively. This allows us to define two scalar densities of edge and screw dislocations of the corresponding slip system
\begin{equation}
\label{eq:2.21}
\begin{split}
\rho^\gta _\perp =\frac{|\mathbf{b}^\gta _{r\perp}|}{b} =\frac{1}{b}|s^\gta _\nu \partial _{\nu ^\gta }\beta^\gta | =\frac{1}{b}\left| \frac{l^\gta _{;1}(\beta ^\gta _{;1}l^\gta _{;1}+\beta ^\gta _{;2}l^\gta _{;2})}{(l^\gta _{;1})^2+(l^\gta _{;2})^2}\right| , 
\\
\rho^\gta _\parallel =\frac{|\mathbf{b}^\gta _{r\parallel }|}{b} =\frac{1}{b}|s^\gta _\tau \partial _{\nu ^\gta }\beta^\gta |=\frac{1}{b}\left| \frac{l^\gta _{;2}(\beta ^\gta _{;1}l^\gta _{;1}+\beta ^\gta _{;2}l^\gta _{;2})}{(l^\gta _{;1})^2+(l^\gta _{;2})^2}\right| ,
\end{split}
\end{equation}
We see that the dislocation densities $\rho^\gta _\perp $ and $\rho^\gta _\parallel $ depend only on the partial derivatives $\beta ^\gta _{;\alpha }$ and $l^\gta _{;\alpha }$, $\alpha =1,2$. For the moving dislocations we allow the loop functions to depend on time $t$ such that the level curves
\begin{equation*}
l^\gta (\xi ^\gta_1,\xi ^\gta _2,c _3,t)=c,
\end{equation*}
with $c_3$ and $c$ being constants, coincide with the dislocation lines during their motion. Similar to the single slip we introduce the normal velocities of dislocation lines as follows
\begin{equation*}
v^\gta _\nu =\mathbf{v}^\gta \cdot \bfnu ^\gta =-\frac{\dot{l}^\gta }{\sqrt{(l^\gta _{;1})^2+(l^\gta _{;2})^2}},
\end{equation*}
with $\dot{l}=l_{,t}$. These kinematic quantities will be used in the model with dissipation.

From the above discussion of kinematics we see that a single crystal with $n$ active slip systems is a generalized continuum with $3+2n$ degrees of freedom at each point: $\mathbf{y}(\mathbf{x})$, $\beta ^\gta (\mathbf{x})$, and $l^\gta (\bfxi ^\gta (\mathbf{x}))$, $\gta =1,\ldots ,n$. We require that the free energy per unit volume of crystal (assumed as macroscopically homogeneous) must be a function of $\mathbf{C}^e=\mathbf{F}^{eT}\cdot \mathbf{F}^e$ (where $\mathbf{F}^e=\mathbf{F}\cdot \mathbf{F}^{p-1}$), $\rho^\gta _\perp $, and $\rho^\gta _\parallel $ 
\begin{equation*}
\psi = \psi (\mathbf{C}^e,\rho^\gta _\perp ,\rho^\gta _\parallel ).
\end{equation*}
Under the same loading condition as for the crystal with single slip we write down the energy functional
\begin{equation}
\label{eq:2.22}
I[\mathbf{y}(\mathbf{x}),\beta ^\gta (\mathbf{x}),l^\gta (\bfxi ^\gta (\mathbf{x}))]=\int_{\mathcal{V}}w(\mathbf{F},\beta ^\gta ,\nabla \beta ^\gta ,\nabla l^\gta )\, dx-\int_{\partial _s} \mathbf{t} \cdot \mathbf{y}\, da,
\end{equation}
where
\begin{equation*}
w(\mathbf{F},\beta ^\gta ,\nabla \beta ^\gta ,\nabla l^\gta )=\psi (\mathbf{C}^e,\rho^\gta _\perp ,\rho^\gta _\parallel ).
\end{equation*}
Provided the resistance to the dislocation motion is negligibly small, we formulate the following variational principle for single crystals with $n$ active slip systems: the true placement vector $\check{\mathbf{y}}(\mathbf{x})$, the true plastic slips $\check{\beta }^\gta (\mathbf{x})$, and the true loop functions $\check{l}^\gta (\bfxi ^\gta (\mathbf{x}))$ in the {\it final} equilibrium state of deformation minimize energy functional \eqref{eq:2.22} among all continuously differentiable fields $\mathbf{y}(\mathbf{x})$, $\beta ^\gta (\mathbf{x})$, and $l^\gta (\bfxi ^\gta (\mathbf{x}))$ satisfying the constraints  
\begin{equation}\label{eq:2.23}
\mathbf{y}(\mathbf{x})=\mathbf{x}+\mathbf{u}_0(\mathbf{x}),\quad \beta ^\gta (\mathbf{x})=0, \quad l^\gta (\mathbf{x})=0 \quad \text{at $\partial _k$},
\end{equation}

Applying the same calculus of variation and taking into account the arbitrariness of the variations of $\mathbf{y}(\mathbf{x})$, $\beta (\mathbf{x})$, and $l(\bfxi ^\gta (\mathbf{x}))$ in $\mathcal{V}$ as well as at $\partial _s$, one can show that the minimizer must satisfy in $\mathcal{V}$ the equilibrium equations
\begin{equation}\label{eq:2.25}
\mathbf{P}\cdot \nabla =0,
\quad
-w_\beta ^\gta +\nabla \cdot w_{\nabla \beta ^\gta }=0, \quad \nabla \cdot w_{\nabla l^\gta } =0,
\end{equation} 
subjected to the kinematic boundary conditions \eqref{eq:2.23} at $\partial _k$ and the following natural boundary conditions 
\begin{equation}\label{eq:2.26}
\mathbf{P}\cdot \mathbf{n}=\mathbf{t},
\quad
w_{\nabla \beta ^\gta }\cdot \mathbf{n}=0, \quad w_{\nabla l^\gta }\cdot \mathbf{n}=0 \quad \text{at $\partial _s$}.
\end{equation}
The constitutive equations for $\mathbf{P}=w_{\mathbf{F}}$, $-w_{\beta ^\gta }$, $w_{\nabla \beta ^\gta }$, and $w_{\nabla l^\gta }$ can easily be obtained from the above free energy density by standard differentiation. For the first Piola-Kirchhoff stress tensor and the Schmid stresses we have 
\begin{equation}\label{eq:2.27}
\mathbf{P}=w_{\mathbf{F}}=2\mathbf{F}^e\cdot \psi _{\mathbf{C}^e} \cdot \mathbf{F}^{p-T}.
\end{equation}
\begin{equation}\label{eq:2.28}
\tau ^\gta_r=-w_{\beta ^\gta }=2\mathbf{s}^\gta \cdot \mathbf{F}^{p-T}\cdot \mathbf{C}^e \cdot \psi _{\mathbf{C}^e}\cdot \mathbf{m}^\gta .
\end{equation}
Likewise, from \eqref{eq:2.21} follows
\begin{equation}
\label{eq:2.29}
w_{\nabla \beta ^\gta }=\frac{1}{b}[\psi _{\rho^\gta _\perp }\sign (s^\gta _\nu \partial _{\nu ^\gta } \beta ^\gta )s^\gta _\nu +\psi _{\rho^\gta _\parallel }\sign (s^\gta _\tau \partial _{\nu ^\gta } \beta ^\gta )s^\gta _\tau ] \bfnu ^\gta .
\end{equation}
Thus, the vectors $w_{\nabla \beta ^\gta }$ are two-dimensional. Finally, for $w_{\nabla l^\gta }$ we have 
\begin{equation}
\label{eq:2.30}
w_{\nabla l^\gta }=\left( \psi _{\rho _\perp ^\gta }\frac{\partial \rho _\perp ^\gta }{\partial l^\gta _{;1}}+\psi _{\rho _\parallel ^\gta }\frac{\partial \rho _\parallel ^\gta }{\partial l^\gta _{;1}}\right)\mathbf{s}^\gta +\left( \psi _{\rho _\perp ^\gta }\frac{\partial \rho _\perp ^\gta }{\partial l^\gta _{;2}}+\psi _{\rho _\parallel ^\gta }\frac{\partial \rho _\parallel ^\gta }{\partial l^\gta _{;2}}\right)\mathbf{p}^\gta .
\end{equation}
Differentiating formulas \eqref{eq:2.21} for the dislocation densities $\rho^\gta _\perp $ and $\rho^\gta _\parallel $ with respect to $l^\gta _{;1}$ and $l^\gta _{;2}$, we get
\begin{align*}
\frac{\partial \rho _\perp ^\gta }{\partial l^\gta _{;1}}&=\frac{1}{b} \sign (s^\gta _\nu \partial _{\nu ^\gta }\beta ^\gta )\left( -\frac{2(l^\gta _{;1})^2(\beta ^\gta _{;1}l^\gta _{;1}+\beta ^\gta _{;2}l^\gta _{;2})}{((l^\gta _{;1})^2+(l^\gta _{;2})^2)^2}+\frac{2\beta ^\gta _{;1}l^\gta _{;1}+\beta ^\gta _{;2}l^\gta _{;2}}{(l^\gta _{;1})^2+(l^\gta _{;2})^2} \right) ,
\\
\frac{\partial \rho _\parallel ^\gta }{\partial l^\gta _{;1}}&=\frac{1}{b}\sign (s^\gta _\tau \partial _{\nu ^\gta }\beta ^\gta )\left( -\frac{2l^\gta _{;1}l^\gta _{;2}(\beta ^\gta _{;1}l^\gta _{;1}+\beta ^\gta _{;2}l^\gta _{;2})}{((l^\gta _{;1})^2+(l^\gta _{;2})^2)^2}+\frac{\beta ^\gta _{;1}l^\gta _{;2}}{(l^\gta _{;1})^2+(l^\gta _{;2})^2} \right) ,
\\
\frac{\partial \rho _\perp ^\gta }{\partial l^\gta _{;2}}&=\frac{1}{b}\sign (s^\gta _\nu \partial _{\nu ^\gta }\beta ^\gta )\left( -\frac{2l^\gta _{;1}l^\gta _{;2}(\beta ^\gta _{;1}l^\gta _{;1}+\beta ^\gta _{;2}l^\gta _{;2})}{((l^\gta _{;1})^2+(l^\gta _{;2})^2)^2}+\frac{\beta ^\gta _{;2}l^\gta _{;1}}{(l^\gta _{;1})^2+(l^\gta _{;2})^2} \right) ,
\\
\frac{\partial \rho _\parallel ^\gta }{\partial l^\gta _{;1}}&=\frac{1}{b} \sign (s^\gta _\tau \partial _{\nu ^\gta }\beta ^\gta )\left( -\frac{2(l^\gta _{;2})^2(\beta ^\gta _{;1}l^\gta _{;1}+\beta ^\gta _{;2}l^\gta _{;2})}{((l^\gta _{;1})^2+(l^\gta _{;2})^2)^2}+\frac{\beta ^\gta _{;1}l^\gta _{;1}+2\beta ^\gta _{;2}l^\gta _{;2}}{(l^\gta _{;1})^2+(l^\gta _{;2})^2} \right) .
\end{align*}
Substituting the constitutive equations \eqref{eq:2.27}-\eqref{eq:2.30} into \eqref{eq:2.25}-\eqref{eq:2.26} we get the completely new system of equations and boundary conditions which, together with \eqref{eq:2.23}, enables one to determine $\check{\mathbf{y}}(\mathbf{x})$, $\check{\beta }^\gta (\mathbf{x})$, and $\check{l}^\gta (\mathbf{x})$.

For the case of non-zero resistance to dislocation motion leading to the energy dissipation we take the dissipation function in the form
\begin{equation*}
D=D(\dot{\beta }^\gta ,\dot{l}^\gta ),
\end{equation*}
We require that the true placement $\check{\mathbf{y}}(\mathbf{x},t)$, the true plastic slips $\check{\beta }^\gta (\mathbf{x},t)$, and the true loop functions $\check{l}^\gta (\mathbf{x},t)$ obey the variational equation
\begin{equation}
\label{eq:2.32}
\delta I+\int_{\mathcal{V}} \sum_{\gta =1}^n (\frac{\partial D}{\partial \dot{\beta }^\gta }\delta \beta ^\gta +\frac{\partial D}{\partial \dot{l}^\gta }\delta l^\gta )\, dx=0
\end{equation}
for all variations of admissible fields $\mathbf{y}(\mathbf{x},t)$, $\beta ^\gta (\mathbf{x},t)$, and $l^\gta (\mathbf{x},t)$ satisfying the constraints \eqref{eq:2.23}. It is then easy to show by exactly the same arguments like those used at the end of the previous Section that equation \eqref{eq:2.32} yields
\begin{equation*}
\mathbf{P}\cdot \nabla =0,
\quad
-w_{\beta ^\gta }+\nabla \cdot w_{\nabla \beta ^\gta }=\frac{\partial D}{\partial \dot{\beta }^\gta }, \quad \nabla \cdot w_{\nabla l^\gta } =\frac{\partial D}{\partial \dot{l}^\gta },
\end{equation*}
which are subjected to the boundary conditions \eqref{eq:2.23} and \eqref{eq:2.26}. The constitutive equations remain exactly the same as \eqref{eq:2.27}-\eqref{eq:2.30}.

\section{Small strain theory}

Let us simplify the above theory for small strains. In this case, instead of the placement $\mathbf{y}(\mathbf{x})$ we regards the displacement $\mathbf{u}(\mathbf{x})$ that is related to the former by
\begin{equation*}
\mathbf{u}(\mathbf{x})=\mathbf{y}(\mathbf{x})-\mathbf{x}
\end{equation*}
as the unknown function. Thus, the total compatible deformation is
\begin{equation*}
\mathbf{F}=\frac{\partial \mathbf{y}}{\partial \mathbf{x}} =\mathbf{I}+\mathbf{u}\nabla .
\end{equation*}
We assume that the displacement gradient $\mathbf{u}\nabla $ (called distortion) is small compared with $\mathbf{I}$. Concerning the plastic deformation given by \eqref{eq:2.16} we also assume that the plastic slips $\beta ^\gta $ are much smaller than 1. Using the the multiplicative resolution \eqref{eq:2.1} to express $\mathbf{F}^e$ through $\mathbf{F}$ and $\mathbf{F}^{p-1}$ and neglecting the small nonlinear terms in it, we get
\begin{equation*}
\mathbf{F}^e=\mathbf{F}\cdot \mathbf{F}^{p-1}=\mathbf{I}+\bfbeta ^e
\end{equation*}
where
\begin{equation*}
\bfbeta ^e =\mathbf{u}\nabla -\sum_{\gta =1}^n \beta ^\gta \mathbf{s}^\gta \otimes \mathbf{m}^\gta .
\end{equation*}
The last equation can be interpreted as the additive resolution of the total distortion into the plastic and elastic parts. The total compatible strain tensor field can be obtained from the displacement field according to
\begin{equation*}
\bfvarepsilon =\frac{1}{2}(\mathbf{u}\nabla +\nabla \mathbf{u}).
\end{equation*}
The incompatible plastic strain tensor field is the symmetric part of the plastic distortion field
\begin{equation*}
\bfvarepsilon ^p =\frac{1}{2}(\bfbeta +\bfbeta ^T)=\frac{1}{2}\sum_{\gta =1}^n \beta ^\gta (\mathbf{s}^\gta \otimes \mathbf{m}^\gta +\mathbf{m}^\gta \otimes \mathbf{s}^\gta ).
\end{equation*}
Accordingly, the elastic strain tensor field is equal to
\begin{equation*}
\bfvarepsilon ^e=\bfvarepsilon -\bfvarepsilon ^p.
\end{equation*}

The dislocation densities remain exactly the same as in the finite strain theory. They are given by the formulas \eqref{eq:2.21}. Concerning the free energy density we will assume that it depends on the elastic strain $\bfvarepsilon^e$ and on the dislocation densities $\rho^\gta _\perp $ and $\rho^\gta _\parallel $
\begin{equation*}
\psi = \psi (\bfvarepsilon^e,\rho^\gta _\perp ,\rho^\gta _\parallel ).
\end{equation*}
The energy of crystal containing dislocations reads
\begin{equation}
\label{eq:3.5}
I[\mathbf{u}(\mathbf{x}),\beta (\mathbf{x}),l(\mathbf{x})]=\int_{\mathcal{V}}w(\mathbf{u}\nabla ,\beta ^\gta ,\nabla \beta ^\gta ,\nabla l^\gta )\, dx-\int_{\partial _s} \mathbf{t}\cdot \mathbf{u}\, da,
\end{equation}
where $w(\mathbf{u}\nabla ,\beta ^\gta ,\nabla \beta ^\gta ,\nabla l^\gta )=\psi (\bfvarepsilon^e,\rho_\perp ,\rho_\parallel )$. Provided the resistance to the dislocation motion can be neglected, then the following variational principle is valid for single crystals: the true displacement field $\check{\mathbf{u}}(\mathbf{x})$, the true plastic slips $\check{\beta }^\gta (\mathbf{x})$, and the true loop functions $\check{l}^\gta (\mathbf{x})$ in the {\it final} state of deformation in equilibrium minimize energy functional \eqref{eq:3.5} among all admissible fields satisfying the constraints
\begin{equation}\label{eq:3.6}
\mathbf{u}(\mathbf{x})=\mathbf{u}_0(\mathbf{x}),\quad \beta ^\gta (\mathbf{x})=0,\quad l^\gta (\mathbf{x})=0  \quad \text{at $\partial _k$}.
\end{equation}
 
The standard calculus of variation similar to the previous case leads to the equilibrium equations
\begin{equation}
\label{eq:3.7}
\bfsigma \cdot \nabla =0, \quad w_\beta ^\gta -\nabla \cdot w_{\nabla \beta ^\gta }=0, \quad \nabla \cdot w_{\nabla l^\gta } =0,
\end{equation}
subjected to the kinematic boundary conditions \eqref{eq:3.6} at $\partial _k$ and the following natural boundary conditions 
\begin{equation}
\label{eq:3.8}
\bfsigma \cdot \mathbf{n}=\mathbf{t},\quad w_{\nabla \beta ^\gta }\cdot \mathbf{n}=0, \quad w_{\nabla l^\gta }\cdot \mathbf{n}=0 \quad \text{at $\partial _s$}.
\end{equation}
The constitutive equation for the Cauchy stress tensor becomes
\begin{equation}
\label{eq:3.9}
\bfsigma =\frac{\partial \psi }{\partial \bfvarepsilon ^e}.
\end{equation}
For the Schmid stress we obtain
\begin{equation}
\label{eq:3.10}
\tau ^\gta _r=-w_{\beta ^\gta }=\mathbf{s}^\gta \cdot \bfsigma \cdot \mathbf{m}^\gta .
\end{equation}
The constitutive equations for $w_{\nabla \beta ^\gta }$ and $w_{\nabla l^\gta }$ remain unchanged as compared with \eqref{eq:2.29} and \eqref{eq:2.30}. The new set of governing equations and boundary conditions are obtained by substituting \eqref{eq:3.9}, \eqref{eq:3.10}, \eqref{eq:2.29}, and \eqref{eq:2.30} into the equilibrium equations \eqref{eq:3.7} and boundary conditions \eqref{eq:3.8}. Note that even for small strain the system of governing equations remain as a whole nonlinear. 

It is a simple matter to modify the theory for the case of non-zero resistance to dislocation motion and plastic slip leading to the energy dissipation.

\section{Simple shear deformation of a single crystal beam}\label{sec:pldeformation}

\begin{figure}[htb]
	\centering
	\includegraphics[width=8cm]{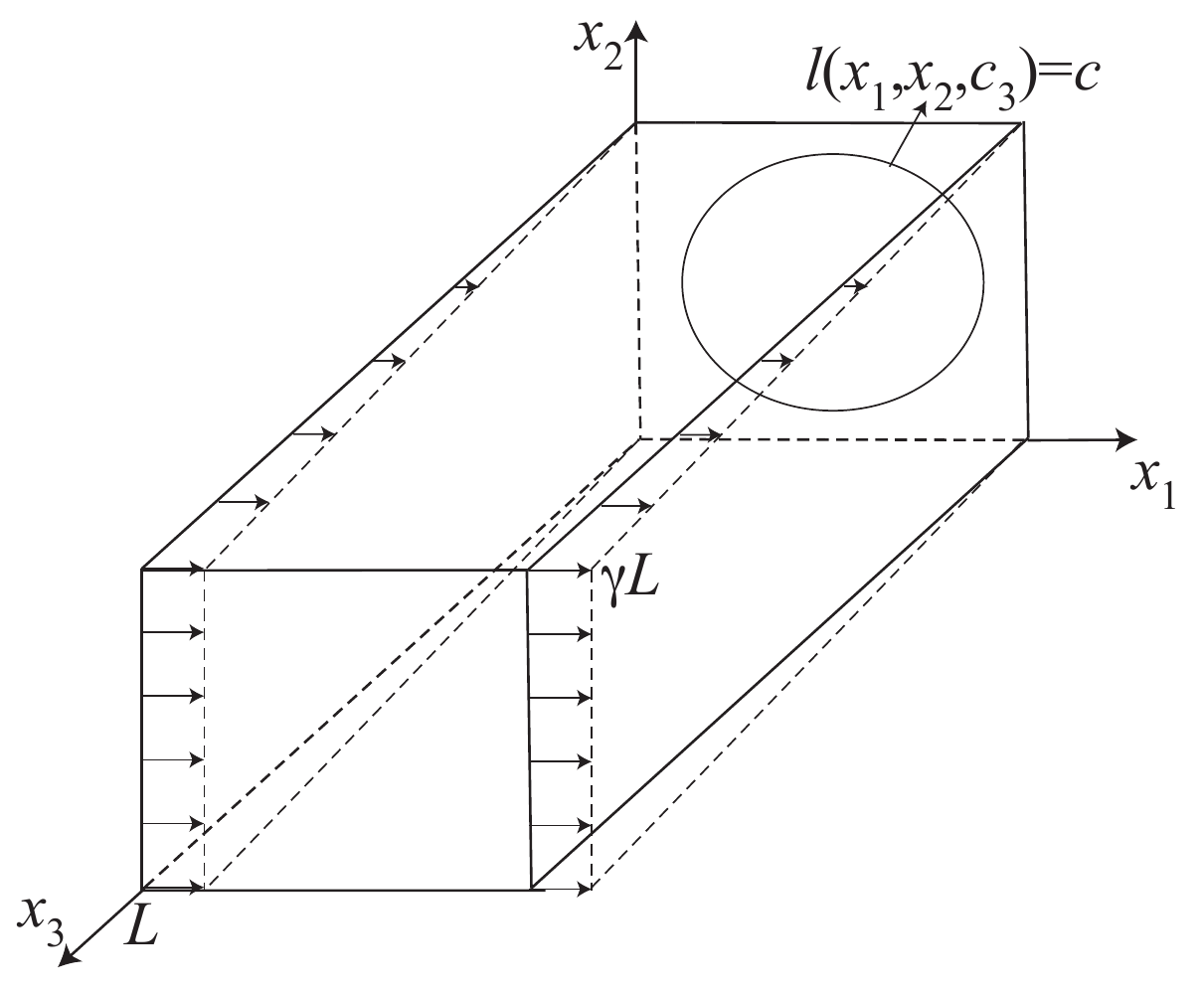}
	\caption{Simple shear deformation of a beam of rectangular cross section}
	\label{fig:plane}
\end{figure}

Let us consider now the simple shear deformation of a single crystal beam having only one active slip system. The crystal occupies in its initial configuration a long cylinder of an arbitrary cross section such that $(x_1,x_2)\in \mathcal{A}$ and $0\le x_3\le L$ (see Fig.~\ref{fig:plane} for the beam of rectangular cross section). As before, the slip system is chosen such that the vectors $\mathbf{s}$, $\mathbf{p}$, and $\mathbf{m}$ coincide with $\mathbf{e}_1$, $\mathbf{e}_2$, and $\mathbf{e}_3$, respectively. We realize the simple shear  deformation by placing this crystal beam in a ``hard'' device with the prescribed displacements at the boundary of the crystal such that
\begin{equation}
\label{eq:5.1}
y_1=x_1+\gamma x_3,\quad y_2=x_2,\quad y_3=x_3.
\end{equation}
We assume that the length of the crystal $L$ is large enough compared with the sizes of the cross section to guarantee the uniform simple shear deformation state for which equation \eqref{eq:5.1} is valid everywhere in the beam. If the overall shear $\gamma $ is sufficiently small, then it is natural to expect that the crystal deforms elastically and the plastic slip as well as the loop function vanish. If this parameter exceeds some critical threshold, then dislocation loops may appear (see one dislocation loop in Fig.~\ref{fig:plane}). Due to the almost translational invariance in $x_3$-direction we may assume that $\beta (\mathbf{x})$ and $l(\mathbf{x})$ depend only on $x_1$ and $x_2$ (except perhaps the neighborhoods of $x_3=0$ and $x_3=L$). 

For simplicity let us consider the small strain theory. Then the only non-zero components of the total strain tensor, under the condition that \eqref{eq:5.1} is valid everywhere, are
\begin{equation*}
\varepsilon _{13}=\varepsilon _{31}=\frac{1}{2}\gamma .
\end{equation*}
Since the plastic distortion tensor $\bfbeta $ has only one non-zero component $\beta _{13}=\beta (x_1,x_2)$, the non-zero components of the elastic strain tensor read
\begin{equation}\label{eq:5.2}
\varepsilon ^e_{13}=\varepsilon ^e_{31}=\frac{1}{2}(\gamma -\beta (x_1,x_2)).
\end{equation} 
With the loop function being $l(x_1,x_2)$ the dislocation densities are given by
\begin{equation}
\label{eq:5.3}
\begin{split}
\rho_\perp  =\frac{1}{b}\left| \frac{l_{,1}(\beta _{,1}l_{,1}+\beta _{,2}l_{,2})}{l_{,1}^2+l_{,2}^2}\right| , 
\\
\rho_\parallel =\frac{1}{b}\left| \frac{l_{,2}(\beta _{,1}l_{,1}+\beta _{,2}l_{,2})}{l_{,1}^2+l_{,2}^2}\right| .
\end{split}
\end{equation}

For the small strain theory we propose the free energy per unit volume of the undeformed crystal in the most simple form
\begin{equation}\label{eq:5.4}
\psi (\bfvarepsilon ^e,\rho _\perp, \rho _\parallel ) = \frac{1}{2}\lambda (\tr \bfvarepsilon ^e)^2 + \mu \tr(\bfvarepsilon^e \cdot \bfvarepsilon ^e) + \frac{1}{2}\mu k_1 \frac{\rho_\perp ^2}{\rho _s^2}+ \frac{1}{2}\mu k_2 \frac{\rho_\parallel ^2}{\rho _s^2} .
\end{equation}
Here $\bfvarepsilon ^e$ is the elastic strain tensor, $\lambda $ and $\mu $ are the Lam\'{e} constants, $k_1$ and $k_2$ are material constants, while $\rho _s$ can be interpreted as the saturated dislocation density. The first two terms in \eqref{eq:5.4} represent the free energy of the crystal due to the macroscopic elastic strain, where we assume that the crystal is elastically isotropic. The last two terms in \eqref{eq:5.4} correspond to the energy of the dislocation network for moderate dislocation densities \citep{Gurtin2002,Gurtin2007}. Note that, for the small or extremely large dislocation densities close to the saturated value, the logarithmic energy proposed by \citet{Berdichevsky06,Berdichevsky06a} turns out to be more appropriate. Substituting formulas \eqref{eq:5.2} and \eqref{eq:5.3} into \eqref{eq:5.4} we obtain
\begin{multline*}
w(\beta ,\nabla \beta ,\nabla l)=\frac{1}{2}\mu (\gamma -\beta )^2+\frac{1}{2}\frac{\mu }{b^2\rho _s^2}\left[ k_1\frac{l_{,1}^2(\beta _{,1}l_{,1}+\beta _{,2}l_{,2})^2}{(l_{,1}^2+l_{,2}^2)^2} \right.
\\
\left. +k_2 \frac{l_{,2}^2(\beta _{,1}l_{,1}+\beta _{,2}l_{,2})^2}{(l_{,1}^2+l_{,2}^2)^2} \right] .
\end{multline*} 
Since in this case the side boundary does not allow dislocations to reach it, we can pose on both functions $\beta (x_1,x_2)$ and $l(x_1,x_2)$ the Dirichlet boundary conditions
\begin{equation}
\label{eq:5.6}
\beta (x_1,x_2)=0,\quad l(x_1,x_2)=0 \quad \text{for $(x_1,x_2)\in \partial \mathcal{A}$},
\end{equation}
The variational problem reduces to minimizing the two-dimensional functional
\begin{equation}
\label{eq:5.7}
I[\beta (x_1,x_2),l(x_1,x_2)]=L\int_{\mathcal{A}} w(\beta ,\nabla \beta ,\nabla l)dx_1dx_2
\end{equation}
among all admissible functions $\beta (x_1,x_2)$ and $l(x_1,x_2)$ satisfying the boundary conditions \eqref{eq:5.6}. 

It is convenient to simplify the functional and minimize it in the dimensionless form. Introducing the dimensionless variables and quantity
\begin{equation*}
\bar{x}_1=b\rho _sx_1, \quad \bar{x}_2=b\rho _sx_2,\quad (\bar{x}_1,\bar{x}_2)\in \bar{\mathcal{A}},\quad \bar{I}=\frac{Ib^2\rho _s^2}{\mu L},
\end{equation*}
we can write functional \eqref{eq:5.7} in the form
\begin{equation}
\label{eq:5.8}
I=\frac{1}{2} \int_{\mathcal{A}}\left[ (\gamma -\beta )^2+ k_1 \frac{l_{,1}^2(\beta _{,1}l_{,1}+\beta _{,2}l_{,2})^2}{(l_{,1}^2+l_{,2}^2)^2} + k_2 \frac{l_{,2}^2(\beta _{,1}l_{,1}+\beta _{,2}l_{,2})^2}{(l_{,1}^2+l_{,2}^2)^2} \right] dx_1dx_2,
\end{equation}
where the bar over the quantities are dropped for short. The problem is to minimize functional \eqref{eq:5.8} among all admissible functions $\beta (x_1,x_2)$ and $l(x_1,x_2)$ satisfying Dirichlet boundary conditions \eqref{eq:5.6}. Since the integrand of \eqref{eq:5.8} is positive definite, the existence of the minimizer in this variational problem is guaranteed.

\begin{figure}[htb]
	\centering
	\includegraphics[width=8cm]{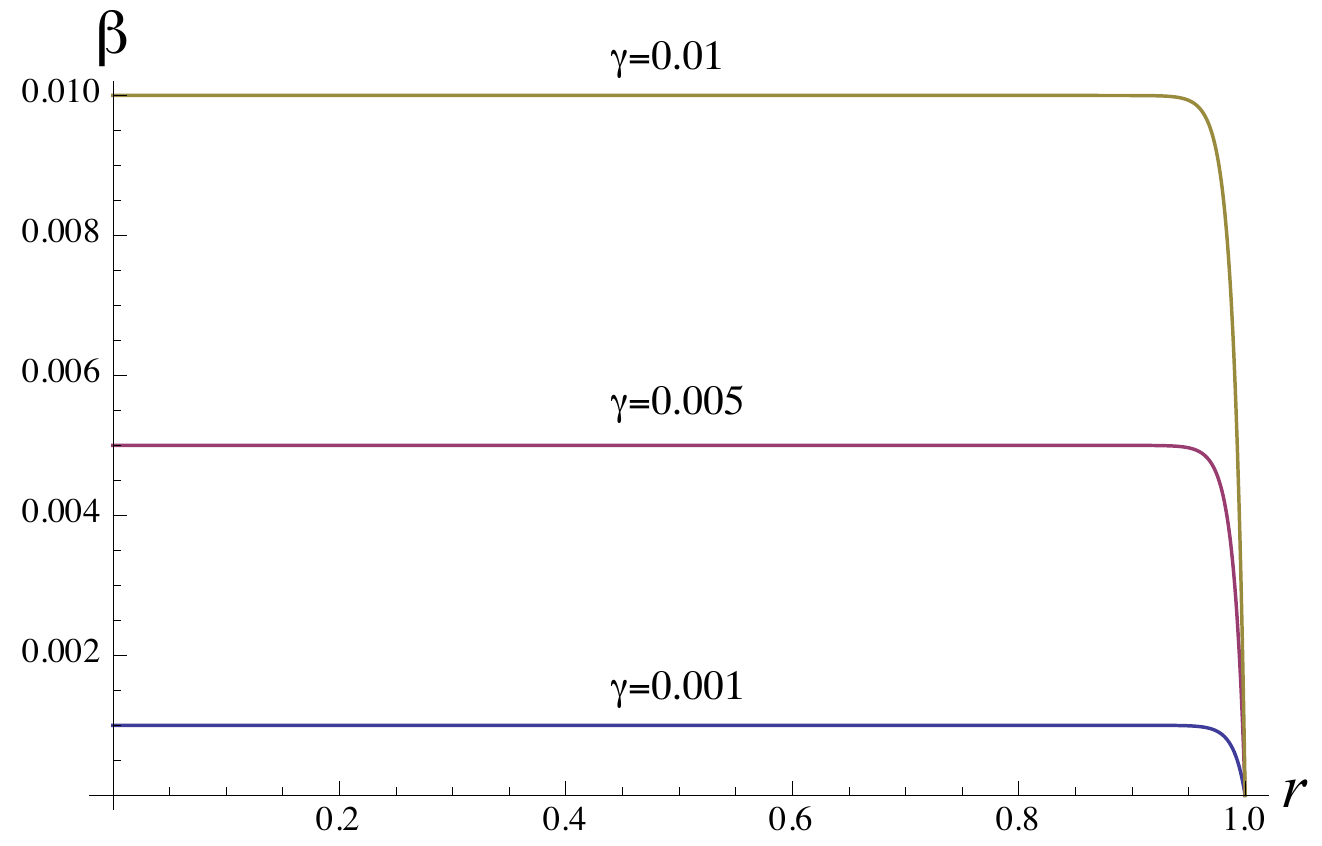}
	\caption{The plastic slip $\beta (r)$: i) $\gamma =0.001$, ii) $\gamma =0.005$, iii) $\gamma =0.01$.}
	\label{fig:sbeta}
\end{figure}

In one special case the problem degenerates and admits an analytical solution. Indeed, if we choose $k_1=k_2=k$ and take $\mathcal{A}$ to be a circular cross section whose boundary is given in the polar coordinates by $r=R$, then due to the symmetry we may assume that both $\beta $ and $l$ are functions of $r=\sqrt{x_1^2+x_2^2}$ only. It is now a simple matter to show that
\begin{equation*}
\frac{l_{,1}^2(\beta _{,1}l_{,1}+\beta _{,2}l_{,2})^2}{(l_{,1}^2+l_{,2}^2)^2} + \frac{l_{,2}^2(\beta _{,1}l_{,1}+\beta _{,2}l_{,2})^2}{(l_{,1}^2+l_{,2}^2)^2}=\beta _{,r}^2.
\end{equation*}
Thus, functional \eqref{eq:5.8} (normalized by $2\pi $) does not depend on $l(r)$ and takes the form
\begin{equation*}
I=\int_0^R [ \frac{1}{2} (\gamma -\beta )^2+\frac{1}{2} k \beta _{,r}^2 ] r dr,
\end{equation*}
which leads to the Euler equation
\begin{equation*}
(\gamma -\beta )+\frac{k}{r}(\beta _{,r}r)_{,r}=0.
\end{equation*}
This is nothing else but the inhomogeneous modified Bessel equation that yields the following solution (regular at $r=0$)
\begin{equation*}
\beta (r)=\gamma +C I_0(r/\sqrt{k}),
\end{equation*}
with $I_0(x)$ being the modified Bessel function of the first kind. The coefficient $C$ must be found from the boundary condition $\beta (R)=0$ giving 
$$C=-\gamma /I_0(R/\sqrt{k}).$$ 
Thus,
\begin{equation}
\label{eq:5.9}
\beta (r)=\gamma [1-I_0(r/\sqrt{k})/I_0(R/\sqrt{k})].
\end{equation}
The plot of solution \eqref{eq:5.9} for different values of $\gamma $ is shown in Fig.~\ref{fig:sbeta}, where we choose the radius such that $\bar{R}=b\rho _sR=1$ and the parameter $k=10^{-4}$. As $\gamma $ increases, the plastic slip increases too. It is seen from this Figure that the plastic slip is nearly constant in the middle of the cross section and changes strongly only in the thin layer in form of ring near the boundary. Since $l$ is a function of $r$, dislocations in form of circles pile up against the boundary of the cross section, leaving the middle of the cross section almost dislocation-free. Since the dislocation loops are circles, they have the purely edge character at $\varphi =0$ and $\varphi =\pi $ and the purely screw character at $\varphi =\pm \pi /2$. For all other angles the dislocation loops have the mixed character. Note that if $k_1\ne k_2$, the strictly axi-symmetric solution is no longer valid because the contributions of the edge and screw components to the energy of dislocation network are not equal. Note also that, if the logarithmic energy is used instead of the quadratic energy, $\beta $ becomes non-zero only if $\gamma >\gamma _c$, so there is a threshold stress for the dislocation nucleation (see \citep{Berdichevsky-Le07}). Besides, the existence of the dislocation-free zone in the middle of the cross-section can be established. 

For an arbitrary cross section and for $k_1\ne k_2$ the problem does not admit exact analytical solution. However, based on the character of solution that changes strongly only in the normal direction to the boundary observed in the previous case, we may use the asymptotic method to find the solution in the thin boundary layer. Take for example the rectangular cross-section $(x_1,x_2)\in (0,W)\times (0,H)$, with $W$ and $H$ being its width and height. In this case let us assume that there are two boundary layers near the left and right boundaries $x_1=0$ and $x_1=W$, and two other boundary layers near the bottom and top boundaries $x_2=0$ and $x_2=H$. Near the boundaries parallel to the $x_2$ axis the derivative with respect to $x_2$ can be neglect as compared to the derivative with respect to $x_1$, while the opposite is true near the boundaries parallel to $x_1$-axis. In the middle of the cross section the plastic slip remains constant. Functional \eqref{eq:5.8} reduces to the sum of four integrals, and all of them do not depend on $l$. Consider for instance the integral along the left boundary layer
\begin{equation*}
I_1[\beta (x_1)]=\int_0^\lambda [ \frac{1}{2} (\gamma -\beta )^2+\frac{1}{2}k_1\beta _{,1}^2]dx_1,
\end{equation*} 
with $\lambda $ being still an unknown length. We omit here the integration over $x_2$ (because in this boundary layer it plays just the role of a parameter) and try to find the minimum among $\beta $. The standard variational calculus leads to the Euler equation
\begin{equation}
\label{eq:5.11}
\gamma -\beta +k_1\beta _{,11}=0
\end{equation} 
which must be subjected to the boundary conditions
\begin{equation}
\label{eq:5.12}
\beta (0)=0.
\end{equation}
The solution of \eqref{eq:5.11} and \eqref{eq:5.12} that does not grow exponentially as $x_1\to \infty $ reads
\begin{equation*}
\beta (x_1)=\gamma (1-e^{-x_1/\sqrt{k_1}}).
\end{equation*}
Due to the mirror symmetry of the problem, the plastic slip in the boundary layer near $x_1=W$ must be 
\begin{equation*}
\beta (x)=\gamma (1-e^{-(W-x_1)/\sqrt{k_1}}).
\end{equation*}
Similarly, the solution in the boundary layers parallel to the $x_1$ axis equals
\begin{equation}\label{eq:5.14}
\beta (x_2)=\begin{cases}
   \gamma (1-e^{-x_2/\sqrt{k_2}})   & \text{near $x_2=0$}, \\
    \gamma (1-e^{-(H-x_2)/\sqrt{k_2}})   & \text{near $x_2=H$}.
\end{cases}
\end{equation}
Thus, the width of the boundary layers parallel to the $x_1$ axis must be of the order $\sqrt{k_2}$, while that of the boundary layers parallel to the $x_2$-axis must be of the order $\sqrt{k_1}$. Since $l$ depends only on the normal coordinate to the boundary, the dislocation lines must be parallel to the boundary of the cross section except at four corners where the asymptotic solution becomes no longer valid (the numerical solution should lead to a smoothing of the corners of dislocation loops). We see that near the vertical boundaries the dislocation loops have the edge character, while near the horizontal boundaries they have the screw character. This agrees well with the widths of the boundary layers determined by the corresponding contributions of the edge and screw components to the energy of the dislocation network.

For very thin rectangular cross section with $H\ll W$ we may neglect the influence of the edges near $x_1= 0$ and $x_1=W$ by considering the dislocation network in the central part of the beam. In this case we will have only screw dislocations which pile up against two obstacle at $x_2=0$ and $x_2=H$. The solution \eqref{eq:5.14} reduces to that found in \citep{Berdichevsky-Le07}.

\section{Conclusion}

In this paper we have developed the nonlinear CDT for crystals containing curved dislocations based on the LEDS-hypothesis. The completely new set of equilibrium equations, boundary conditions and constitutive equations have been derived from the principle of minimum free energy. As the outcome, we have obtained the system of strongly nonlinear partial differential equations for the placement, the plastic slip, and the loop function. In the case of non-vanishing resistance to dislocation motion we have derived the governing equations from the variational equation that takes the dissipation into account. We have extended the theory to the case of multiple slip and simplified it for small strains. The application of the theory has been illustrated on the problem of single crystal beam having one slip system and deforming in simple shear. Under the simplified assumption $k_1=k_2$ the analytical solution of this problem has been found for the circular cross section. For arbitrary cross sections the problem has been solved by the asymptotic method. We have shown that the asymptotic solution found for the rectangular cross section reduces to the well-known solution in \citep{Berdichevsky-Le07} if it is thin and long. 

\bigskip
\noindent {\it Acknowledgments}

The financial support by the German Science Foundation (DFG) through the research projects LE 1216/4-2 and GP01-G within the Collaborative Research Center 692 (SFB692) is gratefully acknowledged.

\end{document}